\documentstyle[12pt]{article}
\newcommand{\hh}{\tilde{h}}
\newcommand{\be}{\begin{equation}}
\newcommand{\ee}{\end{equation}}
\newcommand{\bea}{\begin{eqnarray}}
\newcommand{\eea}{\end{eqnarray}}
\newcommand{\cH}{{\cal H}}
\newcommand{\cK}{{\cal K}}
\newcommand{\prt}{\partial}
\newcommand{\clp}{\mbox{\Large $[$}}
\newcommand{\crp}{\mbox{\Large $]$}}
\newcommand{\clb}{\mbox{\Large $($}}
\newcommand{\crb}{\mbox{\Large $)$}}
\title{ The possibility of a metal insulator transition in \\
antidot arrays induced by an external driving}
\author{A. Iomin and S. Fishman  \\
Department of Physics, Technion, Haifa 32000, Israel }
\begin{document}
\textheight 200mm
\topmargin -0.5cm
\parindent 0mm
\parskip 1em
\reversemarginpar
\begin{titlepage}
\maketitle
\begin{abstract}
It is shown that a family of models associated with the kicked Harper model
is relevant for cyclotron resonance experiments in an antidot array.
For this purpose a simplified model for electronic motion in a related
model system in presence of a magnetic field and an AC electric field
is developed. In the limit of strong magnetic field it reduces to a model
similar to the kicked Harper model. This model is studied numerically
and is found to be extremely sensitive to the strength of the electric 
field. In particular, as the strength of the electric field is varied
a metal -- insulator transition may be found. The experimental conditions
required for this transition are discussed.

\end{abstract}
\thispagestyle{empty}
\end{titlepage}

Transport phenomena of a two dimensional electron gas  embedded in a periodic
potential (antidot array) attracted much attention in theoretical and 
experimental studies \cite{wlr,fgkp}. Investigations of magnetotransport 
\cite{fgkp,fghk} and photo--conductivity \cite{vas} for {\rm GaAs} 
heterostructures show that the electronic dynamics is essentially nonlinear 
with possible transition to a chaotic regime \cite{wlr,fgkp}. Exploration of 
some  chaotic systems illuminates a deep connection between the electron 
transport phenomena and quantum and classical chaos \cite{wlr,fgkp,kkg}. 
Specifically, the kicked Harper model (KHM) is an example of a system which 
is chaotic in the classical limit. It has been introduced theoretically in 
the field of quantum chaos because of its interesting spectral and transport 
phenomena. It is defined by the Hamiltonian 
\begin{equation}
\cH^{KH}=L_H\cos p+K_H\cos q\sum_n\delta(t-n).
\label{eq:i1}
\end{equation}
The model and its variants appear naturally for the kicked harmonic
oscillator \cite{zasl1,dana33}. It is of specific interest since it does
not follow the KAM picture. Its dynamics corresponds to kicks combined with
rotations of the four-fold symmetry.
This system exhibits chaotic motion in a region that increases with $ L_H $
and $ K_H $. The model was subject to extensive theoretical studies
\cite{fgkp},\cite{zasl1}--\cite{dana1}. For some regimes of parameters
its spectrum of quasi-energies is similar in nature to the energy
spectrum of the Harper model \cite{fgkp,geis2,geis3,skg}. It exhibits classical
and quantum diffusion as well as localization and anomalous diffusion
and even ballistic motion \cite{zasl1}--\cite{dana1}. The motivation
for the explorations of the kicked Harper model in the field of quantum chaos
was so far
mainly theoretical, because of the variety of interesting phenomena that were
found. Since the system  can be modeled
approximately by the kicked harmonic oscillator it can be realized
experimentally \cite{dana44}.
Nevertheless, it has been shown recently that this system 
and models associated with KHM may be realized in experiments on cyclotron 
resonance \cite{ifprl}. It was proposed there that the KHM can be used to 
model some aspects of electronic motion in antidot arrays.
In this paper we show that a family of models associated with
KHM is relevant for cyclotron resonance experiments \cite{vas} in antidot 
arrays for some conditions. 

The one particle Hamiltonian describing electrons in external fields and
in a periodic potential is
\begin{equation}
H=\frac{1}{2m^*}\clb{\bf p}-\frac{e}{c}{\bf A}({\bf r},t)\crb^2+
V({\bf r}),
\label{eq:f1}
\end{equation}
where $ {\bf p}=(p_x,p_y) $ is the two dimensional momentum of an electron
with effective mass $ m^* $ and charge $ e $, $ {\bf A}({\bf r},t)=
(\frac{c}{\nu}E_x\sin\nu t, xB-\frac{c}{\nu}E_y\cos\nu t,0) $ is the vector
potential in such a gauge that takes into account a constant magnetic
field in the $z$--direction $ {\bf B}=(0,0, B) $ and an alternating electric
field $ {\bf E}=(-E_x\cos\nu t,\,E_y\sin\nu t) $ with a frequency $ \nu $,
while $ V({\bf r})=V(x,y) $ is a two dimensional periodic potential modeling 
the antidot structure. For simplicity the potential
\be
V({\bf r})=V_x\cos\frac{2\pi x}{a}+V_y\cos\frac{2\pi y}{b},
\label{eq:f2}
\ee
where $ a,~ b $ are periods in the $ x $ and $ y $ directions,
will be considered, although it should be much sharper and 
contain more Fourier components for the antidot lattice 
\cite{fgkp}-\cite{kkg}. 
The equations of motion for (\ref{eq:f1}) are
\bea
\dot{p}_x=\omega^*(p_y-\omega^*m^*x+\frac{e}{\nu}E_y\cos\nu t)-
\frac{\prt V(x,y)}{\prt x}, ~~~
\dot{x}=\frac{p_x}{m^*}-\frac{e}{m^*\nu}E_x\sin\nu t, \nonumber \\
\dot{p}_y=-\frac{\prt V}{\prt y}, ~~~
\dot{y}=\frac{p_y}{m^*}-\omega^*x+\frac{e}{m^*\nu}E_y\cos\nu t,
\label{eq:f3}
\eea
where $\omega^*=\frac{eB}{m^*c} $ is the electronic cyclotron frequency. 
The following change of variables 
$ y=\tilde{y}+y_0\sin\nu t $ with $ y_0=\frac{eE_y}{m^*\nu^2} $ is useful.
The equations of motion for the $ y $ component in terms of this variable 
are
\be
\dot{\tilde{y}}=\frac{p_y}{m^*}-\omega^*x,~~~\dot{p}_y=-\frac{\prt V(x,y)}
{\prt \tilde{y}}.
\label{eq:f4}
\ee
The fact that $ \frac{\prt}{\prt y}=\frac{\prt}{\prt\tilde{y}} $ was used. 
Analogously, we make change
of variables $ p_x=\tilde{p}_x+\frac{e\omega^*}{\nu^2}E_y\sin\nu t $,
leading to
\be
\dot{\tilde{p}}_x=\omega^*(p_y-x\omega^*m^*)-\frac{\prt V(x,y)}{\prt x},
~~~~
\dot{x}=\frac{1}{m^*}(\tilde{p}_x+\frac{\omega^* e}{\nu^2}E_y\sin\nu t-
\frac{e}{\nu}E_x\sin\nu t).
\label{eq:f5}
\ee
If the amplitudes $ B,~E_x,~E_y $ satisfy 
\be
 \frac{\omega^*}{\nu}E_y=E_x ,
\label{eq:ad1}
\ee 
the equation of motion for $ x $ is particularly simple and reduces to
$ \dot{x}=\tilde{p}_x/m^* $. 
The equations of motion (\ref{eq:f4}),(\ref{eq:f5}) result from the
effective Hamiltonian
\bea
\cH=\frac{1}{2m^*}[p_x^2+(p_y-\omega^*m^*x)^2]+V_x\cos\frac{2\pi x}{a}+
V_y\cos\clb\frac{2\pi}{b}(y+y_0\sin\nu t)\crb \nonumber \\
\equiv \cH_0(p_x,p_y,x)+V(x,y,t).
\label{eq:f6}
\eea
The tilde is omitted in (\ref{eq:f6}) and in what follows for the sake 
of simplicity of notation.
Here $ \cH_0=\cH_0(p_x,p_y,x) $ corresponds to the integrable system  
describing the cyclotron motion, while $ V=V(x,y,t) $ is the perturbation
leading to classical chaos.

As the Hamiltonian $ \cH $ is periodic in time with the period 
$T=\frac{2\pi}{\nu} $ the following quantum mechanical analysis will be
carried out in the framework of the Floquet theory. 
The eigenvalue problem of the Floquet operator
\be
\hat{F}=-i\hbar\frac{\prt}{\prt t}+\hat{\cH}_0+\hat{V}=\hat{F}_0+\hat{V}
\label{eq:f7}
\ee

is considered. An unperturbed basis of $ \hat{F}_0 $ is
\be
|j,s,u>=\frac{\nu}{\sqrt{2\pi}}e^{-i j\nu t}|s,u>.
\label{eq:f8}
\ee
The wave function $ |s,u> $ in the coordinate representation is
\cite{skg,pg}:
\be
<x,y|s,u>=\frac{1}{\sqrt{2\pi lb}}e^{i yu/b}\psi_s(\frac{x}{l}-
\frac{u l}{b}),
\label{eq:f9}
\ee
where $ l=\sqrt{\hbar c/eB} $ is the magnetic length, 
while $ \psi_s(z)=
\clp\exp(-z^2/2)/\sqrt{\sqrt{\pi}2^ss!}\crp H_s(z) $ is a   
parabolic cylinder function and $ H_s(z) $ is the $s$-th Hermite polynomial. 
The matrix of $ \hat{F}_0 $ in this basis is diagonal,
\be
<j',s',u'|\hat{F}_0|j,s,u>=[\hbar\omega^*(s+\frac{1}{2})-
\hbar\nu j]\delta_{s,s'}\cdot\delta_{j,j'}\cdot\delta(u-u').
\label{eq:f10}
\ee
The matrix elements of $ \hat{V} $ can be found with the help of the
generating function of the Hermite polynomials and take the form
\bea
<j',s',u'|\hat{V}|j,s,u>=V_xP_{s',s}^{+}(\hh\alpha^{-1})
\cos(\hh u -\frac{\pi}{2}|s'-s|)\cdot\delta_{j,j'}\cdot\delta(u-u') \nonumber \\
+\frac{V_y}{2}[P_{s',s}^{-}(\hh\alpha)J_{j-j'}(\kappa)\delta(u-u'+2\pi)
+(-1)^{s'-s}P_{s',s}^{-}(\hh\alpha)J_{j'-j}(\kappa)\delta(u-u'-2\pi)].
\label{eq:f11}
\eea
where  $ J_n(z) $ is the Bessel function, and 
\be
P_{s',s}^{\pm}(w)=
e^{-\pi w/2}\sqrt{\frac{s!}{s'!}}(2\pi w)^{\frac{|s-s'|}{2}}
L_s^{|s-s'|}(\pm\pi w)
\label{eq:f12}
\ee
for $ s'>s $, while for $ s'<s $ one should interchange $ s' $ and $ s $.
Here $ L_n^k(z) $ are the associated Laguerre polynomials, while
$ \alpha=b/a $ and the effective Planck constant is
$ \hh =\frac{2\pi l^2}{ab}=\frac{\Phi_0}{\Phi} $. The magnetic flux
quantum is $ \Phi_0=\frac{hc}{e} $ and the flux through a unit cell
of the periodic structure is $ \Phi=abB $. 
Because of the periodicity in y the allowed transitions change $ u $
only by integer multiples of $ 2\pi $. This is also clear from
(\ref{eq:f11}). For this reason it is convenient to express $ u $ in the
form $ u=2\pi n +\vartheta $ where $ n $ is an integer while
$ 0<\vartheta<2\pi $. During the evolution $ \vartheta $ is constant.

The eigenstates $ |\lambda,\vartheta> $ of the Floquet operator $ \hat{F} $
satisfying $ \hat{F}|\lambda,\vartheta>=\lambda|\lambda,\vartheta> $
are decomposed into the unperturbed basis states 
\be
|\lambda,\vartheta>=\sum_{n,s,j}c_{n,s}^j|j,s,n,\vartheta>.
\label{eq:f13}
\ee
Projecting on the state $ < j,s,n,\vartheta| $ one finds that
the coefficients $ c_{n,s}^j $ satisfy the following equation:
\bea
\clp\tilde{\lambda}+\nu j\crp c_{n,s}^j=
\tilde{L}\sum_rP_{s,r}^+(\hh\alpha^{-1})
\cos\clp\hh(2\pi n+\vartheta)-\frac{\pi}{2}|s-r|\crp c_{n,r}^j  \nonumber \\
+\frac{\tilde{K}}{2}\sum_{r,j'}[P_{s,r}^-(\hh\alpha)
J_{j'-j}(\kappa)c_{n+1,r}^{j'} +(-1)^{s-r}P_{s,r}^-(\hh\alpha) 
J_{j-j'}(\kappa)c_{n-1,r}^{j'}]
\label{eq:f15}
\eea
The following change of variables was used here: $ 
\nu/\omega^*\rightarrow\hh\nu $
and $ \hh\tilde{\lambda}=[\lambda-\hbar\omega^*(s+\frac{1}{2})]/
[\hbar\omega^*] $, while
$ \tilde{L}=\sqrt{\frac{V_x}{V_y}}\cdot\cK $,  and 
$\tilde{K}=\sqrt{\frac{V_y}{V_x}}\cdot\cK $. 
The parameter $ \hh\cK=\sqrt{V_xV_y}/\hbar\omega^* $ is the strength of 
the coupling of the Landau bands introduced in \cite{skg} for 
a stationary problem of Hall conductance in this system. In the case when 
it is not too large ($ \cK\le 6 $) the coupling between the Landau bands 
is weak \cite{skg}. In the quantum case when $ \hh\alpha<1 $ and 
$ \hh\alpha^{-1}<1 $ we can neglect this interaction, {\rm i.e.} the terms 
with $ s\neq s' $ in (\ref{eq:f15}) can be omitted. 
This can be seen from (\ref{eq:f12}) since the leading term of
$ P_{s',s}(w) $ behaves as $ w^{|s'-s|/2} $. Within this approximation
(\ref{eq:f15}) reduces to:
\be
[\tilde{\lambda}+\nu j] c_n^j=
L\cos[\hh(2\pi n+\vartheta)]c_n^j+
\frac{K}{2}\sum_{j'}J_{j-j'}(\kappa)[(-1)^{j-j'}c_{n+1}^{j'}+c_{n-1}^{j'}],
\label{eq:f16}
\ee
where $ L=\tilde{L}P_{s,s}^+(\hh\alpha^{-1})/P_{s,s}^-(\hh\alpha) $,
$ K=\tilde{K} $, and $ [\tilde{\lambda}+\nu j]/P_{s,s}^-(\hh\alpha)
\rightarrow  [\tilde{\lambda}+\nu j] $,
while the index $ s $ was suppressed in the notation. The Hamiltonian
\be
\hat{H}_1=L\cos\hat{p}+K\cos(\hat{q}+\kappa\sin\nu t),
\label{eq:f18}
\ee
reduces to (\ref{eq:f16}) in the basis $ e^{ikq}e^{-ij\nu t} $,
where $ k=\frac{u}{2\pi}=\frac{2\pi n+\vartheta}{2\pi} $ and 
the canonical variables are $ \hat{q} $ and 
$ \hat{p}=-i2\pi\hh\frac{\prt}{\prt q} $
satisfying $ [\hat{p},\hat{q}]=-i2\pi\hh $.
In the absence of the AC electric field $ \kappa=0 $ and (\ref{eq:f18})
reduces to the Hamiltonian of the Harper model.
For the model of this paper the semi-classical limit $ \hh\rightarrow 0 $
is the limit of strong magnetic field $ \Phi\rightarrow\infty $ (or 
$ B\rightarrow\infty $). The validity of equations (\ref{eq:f16}) and
(\ref{eq:f18}) requires $ \frac{\Phi_0}{\Phi}=\hh<1 $, therefore the 
classical limit is meaningful for (\ref{eq:f18}). This Hamiltonian was 
studied in \cite{ifprl}. It was derived there for a model opposite to the one 
studied here, namely the tight binding model where  the external fields 
can be considered as a perturbation on the lattice potential.

The driving potential $ K\cos(q-\kappa\sin\nu t) $ (related to the one 
of (\ref{eq:f18}) by a trivial shifting of time) is well known in the 
literature and it has been discussed in the context of the description of
dynamical localization
in atomic momentum transfer  \cite{gsz,gss}. This effect has been observed
experimentally following an extension of a theoretical proposal
\cite{Rizen}.
The Hamilton equations for  $ {H}_1 $ of (\ref{eq:f18})
are $ \dot{q}=-L\sin p $ and $ \dot{p}=K\sin\psi(t) $ with $
\psi(t)=q-\kappa\sin\nu t $. For $ \kappa\gg 1 $
the forcing resulting in change of momentum is dominated by
the resonant points \cite{ifprl}, where $ \dot{\psi}=0 $ or
\be
-L\sin p=\dot{q}=\kappa\nu\cos\nu t.
\label{eq:f19}
\ee
Expanding $ \psi(t) $ around the resonant point $ t_r $
and integrating the Hamilton equation for $ p $,
taking into account the fact that this integral
accumulates most of its contribution from a narrow region (of width
$ |\kappa\nu^2\sin \nu t_r|^{-1/2} $) around the resonance, one finds that
the momentum transferred at each resonance is
\be
\Delta p_r^{\pm}=\sqrt{\frac{2\pi}{\kappa\nu^2|\sin\nu t_r|}}K\sin
(\psi_r^{\pm}\pm\frac{\pi}{4}),
\label{eq:for20}
\ee
where $ \psi_r=\psi(t_r) $ and the sign depends on the direction of
crossing of the resonance.

The position of the resonance depends on $ p $. For strong driving so
that $ \kappa\nu\gg L $ for the resonance condition it is required that $ \cos\nu t_r\approx 0 $
or
$ |\sin\nu t_r|\approx 1 $. Consequently the resonant points are $ \nu
t_r\approx\pm\frac{\pi}{2} $ and therefore are approximately equally
spaced in time, and occur at the times: $ \nu t_r^-=-\frac{\pi}{2}+2\pi l $
and $ \nu t_r^+=\frac{\pi}{2}+2\pi l $ ($l$ are integers).
The resulting map is
$ M=M_2\cdot M_1 $ with
\bea
M_1:&~~~~~\left\{\begin{array}{ll}
p_1=&p+K_1\sin(q+\kappa_0) \\
q_1=&q-L_1\sin p_1
\end{array} \right. \nonumber \\
M_2:&~~~~~\left\{\begin{array}{ll}
p_2=&p_1+K_1\sin(q_1-\kappa_0) \\
q_2=&q_1-L_1\sin p_2
\end{array} \right.
\label{eq:f21}
\eea
where $ L_1=\frac{\pi}{\nu}L $ while $ K_1=\sqrt{\frac{2\pi}
{\kappa\nu^2}}K $ and $  \kappa_0 =\kappa-\frac{\pi}{4} $.
First we note that the map is periodic in $ \kappa $ with the period
$ 2\pi $ and therefore it is periodic  in the magnitude of the electric
field.
This map is generated by the Hamiltonian:
\bea
\cH_2&=&L\cos p+\nu K_1\cdot\clp \cos(q+\kappa_0)
\sum_{n=-\infty}^{\infty}\delta(\nu t+\frac{\pi}{2}-2\pi n) \nonumber \\
&+&\cos(q-\kappa_0)\sum_{n=-\infty}^{\infty}\delta(\nu
t-\frac{\pi}{2}-2\pi n)\crp.
\label{eq:f22}
\eea
In the case when $ \kappa_0=0 $ this system corresponds to the KHM.
Numerical analysis shows that the model is extremely sensitive to the parameter
$ \kappa_0 $ in both quantum $ (\hh\sim 1) $ and classical $ (\hh=0) $ 
regimes. In the quantum regime we study numerically the time evolution of 
the variance
\be
var(t)=<\hat{n}^2(t)>-<\hat{n}(t)>^2=\sum_nn^2|f_n(t)|^2
\label{eq:f23}
\ee
and energy spreading over the unperturbed level spectrum as a result of 
the evolution from the initial level occupation $ f_n(0) $ in the momentum 
space. The standard technique of the fast 
Fourier transform is used to evolve the quantum map (\ref{eq:f22}) of the 
wave function over the period $ T $ over which two kicks take place:
$ \Psi(t+T)=\hat{U}\Psi(t) $, where $ \hat{U} $ is the evolution operator 
over the period and $ \Psi(t,q)=\frac{1}{\sqrt{2\pi}}
\sum_{n=-\infty}^{\infty}e^{2\pi i nq}f_n(t) $. The initial distribution 
used is $ f_n(0)=\delta_{n,0} $.
An essential deviation from KHM is found as the parameter $ \kappa $
changes. The parameters $ K_1,~L_1 $ and $ \hh $ are the same 
as in \cite{dima2} and are explicitly specified in the figures.
  
It is found here that when the parameter $ \kappa_0 $ is varied,
localization-delocalization transition takes place, with different
realizations of the delocalization as diffusion, anomalous diffusion 
and ballistic motion.
Such a transition  from diffusive motion for $ \kappa_0=0 $ to ballistic 
one for $ \kappa_0=g\pi $ where $ g=\frac{\sqrt{5}-1}{2} $ is the golden 
mean is shown in Fig.~1.  A localization-delocalization transition is found 
also for $ K_1/L_1<2 $ ( see Fig.~2), such that for KHM ($\kappa_0=0$) there 
is dynamical 
localization of quasi-energies, while ballistic motion is found for 
$ \kappa_0=g\pi $. When $ K_1,L_1\ll 1 $ a slow diffusive process in 
momentum (reported first in Ref.~\cite{dima2})
is observed for KHM in the semi-classical limit. It is suppressed
when the parameter $ \kappa_0 $ increases from zero as shown in Fig.~3
resulting in transport motion in real space.
This transition corresponds to a metal--insulator transition in 
the $y$-direction of the sample. Note the relation between the components of
the electric field (\ref{eq:ad1}) that is crucial for the preference of the
$ y $ direction.
It is enhanced in the limit when $ \hh=0 $.
Numerical iterations of the classical map (\ref{eq:f21}) for small values 
$ K_1 $ and $ L_1 $  with  initial  distribution taken in the vicinity of a 
separatrix are presented in Figs 4 and 5. These calculations demonstrate
that this transport in the $ y $ direction found for finite values of
$ \kappa_0 $  and absent when it vanishes (for the same values of $ K_1 $ 
and $ L_1 $) is of pure classical origin. Therefore it is expected to
be robust against effects of noise.

A model for transport in a two dimensional electron gas embedded
in the periodic super-lattice potential in the presence of external 
fields was studied. The motion corresponds to cyclotron resonance dynamics
with possible metal--insulator transitions induced by the high frequency 
driving. It can be realized experimentally by a 2D periodic andidot
array. The required conditions for experimental realization of this effect 
can be achieved \cite{vas}. For $ m^*\sim 0.1m_e $, $ a,b\sim 10^2\div 4\cdot 10^2 $nm, $ B\sim 
0.07\div0.37 $T, one obtains that $ \omega^*\sim 10^{11}\div 5\cdot 
10^{11} $s$^{-1}$, while $ \hh\sim 0.08\div 6 $.
For $ \nu\sim 10^{11} $s$^{-1}$ and $ E_0\sim 1 $CGSE one obtains
that $ \kappa\sim 300 $ and the condition $ \kappa\gg 1 $ is fulfilled. 
The amplitudes of the periodic potential can satisfy
the condition $ \cK=\frac{m^*ab\sqrt{V_xV_y}}{2\pi\hbar^2}\leq 6 $ for
$ V_x,V_y<4 $meV and $ a\sim b\sim 10^2 $nm.
Reasonable values for the parameters of (\ref{eq:f21}) and (\ref{eq:f22})
are $ K_1=L_1=5 $ that are obtained for $ V_x=0.15 $meV and $ V_y=3 $meV
for $ a=b\sim 10^2 $nm, $ \kappa_0=300 $ and $ \nu=10^{11} $s$^{-1}$.
These were used in numerical calculations of the present paper.
The values of $ K_1 $ and $ L_1 $ can be reduced and also 
numerical calculations for smaller values were performed.

Two metal--insulator transitions induced by the variation of $ \kappa_0 $
are predicted. The first one is  a quantum effect of strong delocalization
in $ p $ corresponding to anomalous (super) diffusion or acceleration
 for $ p $, so that $ <(\hat{p}-p_0)^2>\sim t^{\mu} $ with
$ 1<\mu\leq 2 $ for $ \kappa_0\neq 0 $ and for any ratio between $ K_1 $ 
and $ L_1 $, while it is known that for $ \kappa_0=0 $,
that corresponds to the KHM, super-diffusion takes place only for $ 
K_1>L_1 $. It means that for $ \kappa_0\neq 0 $ an initial wave packet 
spreads in the direction of stronger modulation of the 2D super-lattice 
potential.
So far, in  the one-band structure systems both integrable like the Harper 
model and chaotic like the kicked Harper model ($ \kappa_0=0 $), wave 
packets were 
found to spread in  the direction of the weaker modulation of the potential 
amplitudes. Recently such effect has been studied in the many-band structure 
system with strong interaction between the bands leading to chaos in the 
classical limit \cite{kksg1}. It corresponds to electronic transport in a 
2D periodic 
potential with different modulation amplitudes in $ x $- and $ y $- 
directions that are perpendicular to the magnetic field.
The second one is a classical effect taking place for $ K_1,L_1< 1 $ in the
both classical $ \hh=0 $ and semi-classical limit $ \hh\ll 1 $. It corresponds
to the strong delocalization in the $ q $ direction with anomalous diffusion 
or ballistic motion at any ratio  $ \frac{K_1}{L_1} $ for 
$ \kappa_0\neq 0 $ while for $ \kappa_0=0 $ anomalous diffusion in $ q $ 
takes place only when $ \frac{K_1}{L_1}<1 $.
This effect can be explained by the topological reconstruction of 
phase space due to a variation of $ \kappa_0 $ leading to different 
settling of elliptic and hyperbolic points. An analogous phenomenon has been 
studied for the web map \cite{pr_k} and it has been shown that there 
is a connection between the web structure and features of diffusion in phase 
space \cite{dana1}. For the first case the quantum transition due to 
the non-vanishing $ \kappa_0 $ corresponds to  delocalization in the $ p $ 
direction and localization in $ q $. The second transition corresponds to 
delocalization in $ q $ and localization in $ p $.
Some caution is required when the model studied here is compared to antidot
lattices, where the potential $ V({\bf r}) $ in (\ref{eq:f1}) 
should be replaced
by a sharper function. In this case the equation 
(\ref{eq:f16}) and the 
effective Hamiltonian (\ref{eq:f18}) have more 
complicated form due to
interaction between $ x $ and $ y $ degrees of freedom. 
For  a sufficiently strong magnetic field, so that $ \hh\alpha $ and
$ \hh\alpha^{-1} $ are much smaller than unity, and $ \cK< 6 $
the one Landau band approximation is valid \cite{skg} and
the metal--insulator transitions considered here can be observed.
(This justifies ignoring the  terms with $ s\neq s' $ in (\ref{eq:f11}).)
The result may be modified since the form of the equation corresponding to
(\ref{eq:f18}) may be somewhat more complicated because of the sharpness
of the potential.

This research was supported in part by the Israel Science Foundation,
by the U.S.--Israel Binational Science Foundation (BSF) and by the 
Niedersachsen Ministry of Science (Germany).

\newpage
\section*{Figure Captions}
\begin{description}
\item[Fig.~1]
The quantum occupation probability $ |f_n(t)| $ vs $ n $ and the variance
$ var(t) $ found from integration of (\ref{eq:f22}) for $ K_1=L_1=5 $ 
and $ \hh=\frac{2\pi}{7+g} $ with $ \kappa_0=0 $ (a,b) and $ \kappa_0=\pi g 
$ (c,d). \item[Fig.~2]
Same as Fig.~1 but for $ L_1=4 $ and $ K_2=2 $.
\item[Fig.~3] 
Same as Fig.~1 for $ K_1=L_1=\frac{\pi}{8} $ abd $ \hh=\frac{2\pi}{222+g} $
for $ \kappa_0=0 $ (a,b) and $ \kappa_0=g\pi $ (c,d).
\item[Fig.~4]
Classical evolution of the map (\ref{eq:f21}) for $ K_1=L_1=\frac{\pi}{8} $
and $ \kappa_0=0 $. Eleven trajectories are presented in figure (a) and
folded to the $ 2\pi\times 2\pi $ torus in figure (b). The variances
are presented in (c) and (d). 
\item[Fig.~5]
Same as Fig.~4 but for $ \kappa_0=0.8\cdot g\pi $.
\end{description}
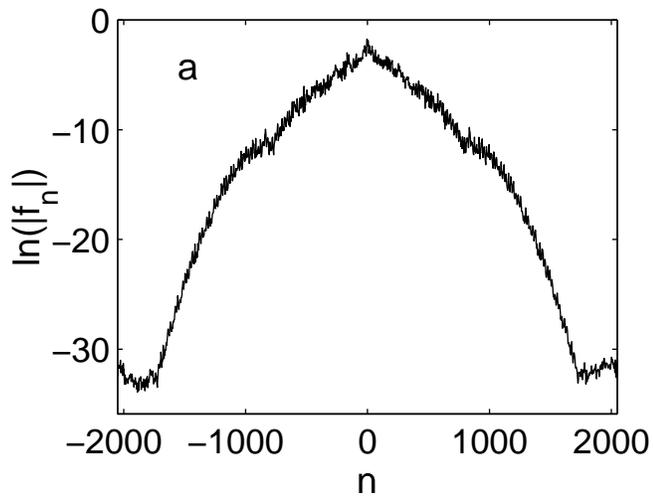
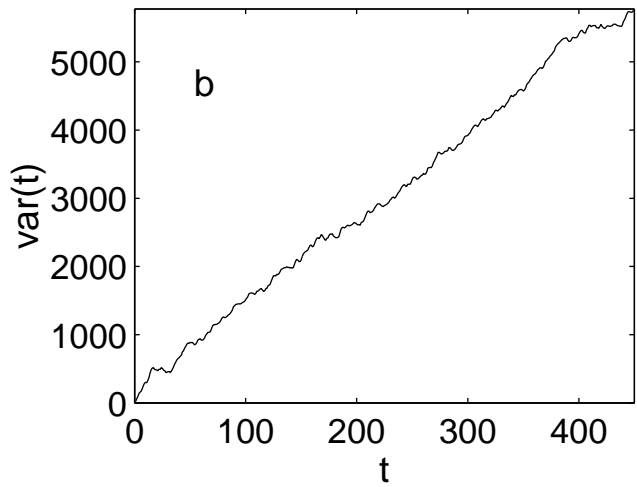
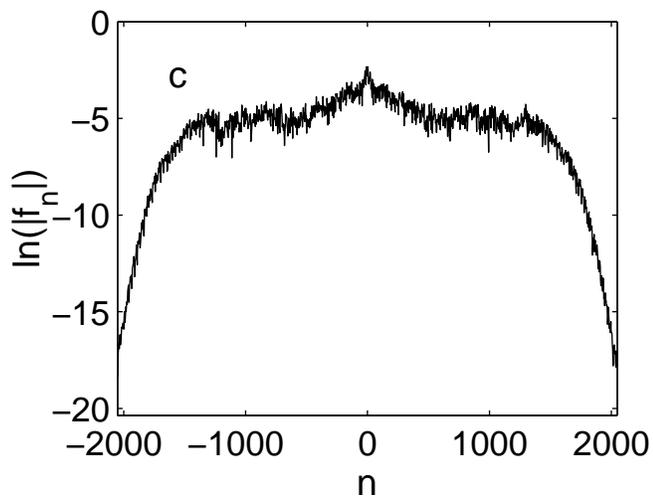
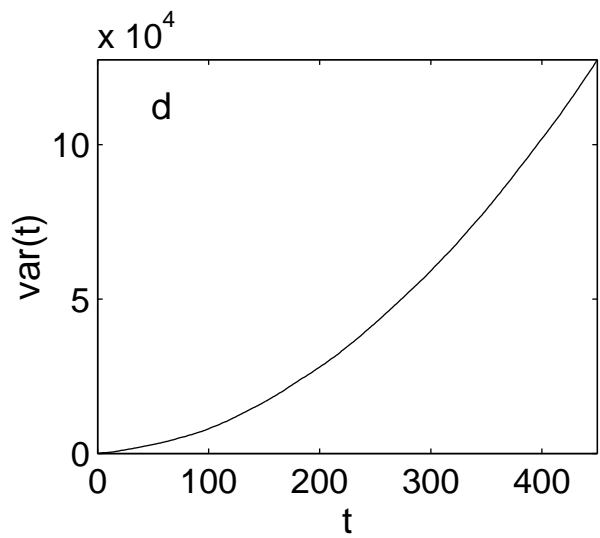
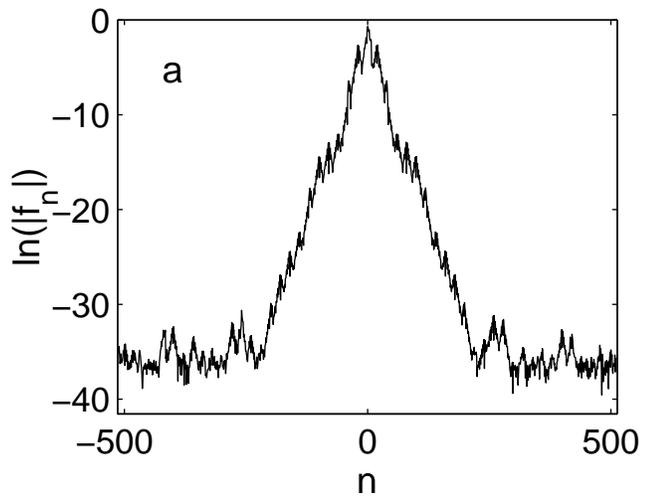
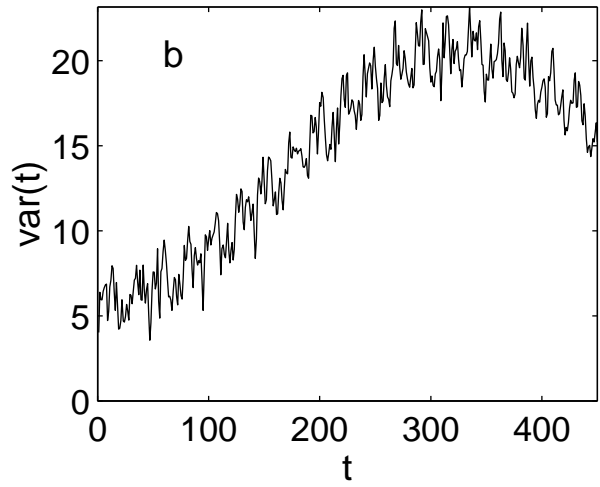
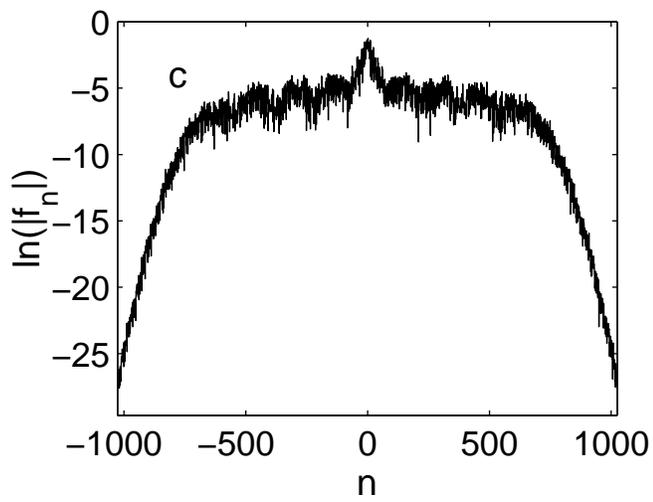
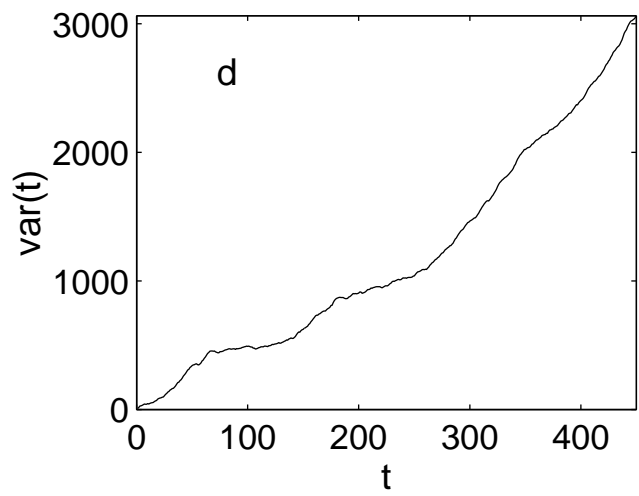
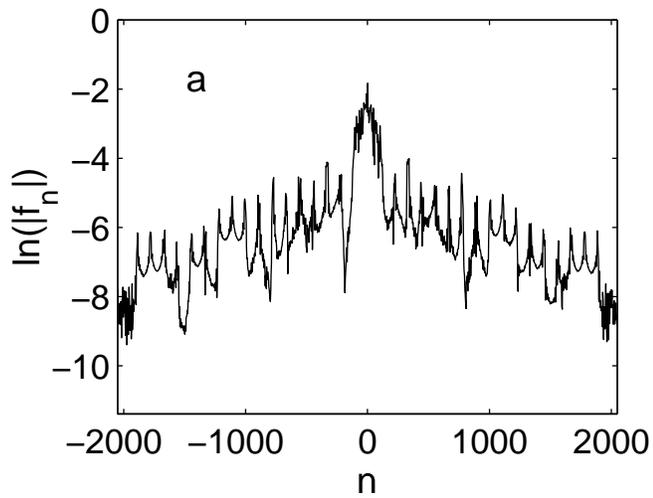
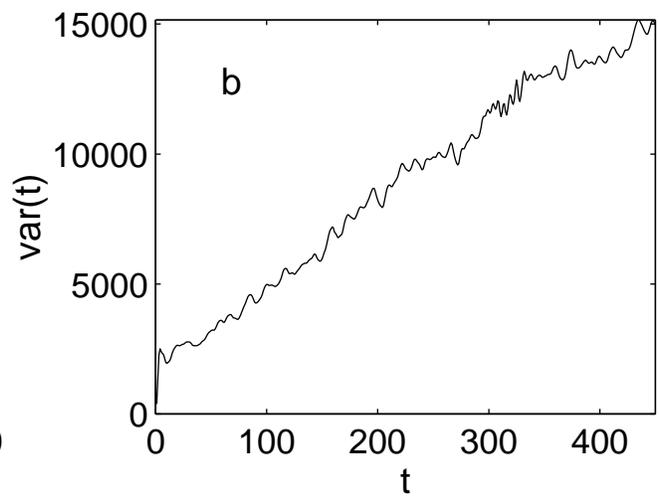
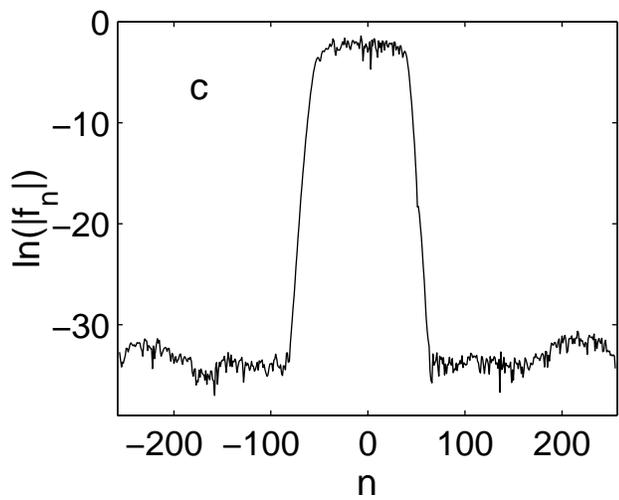
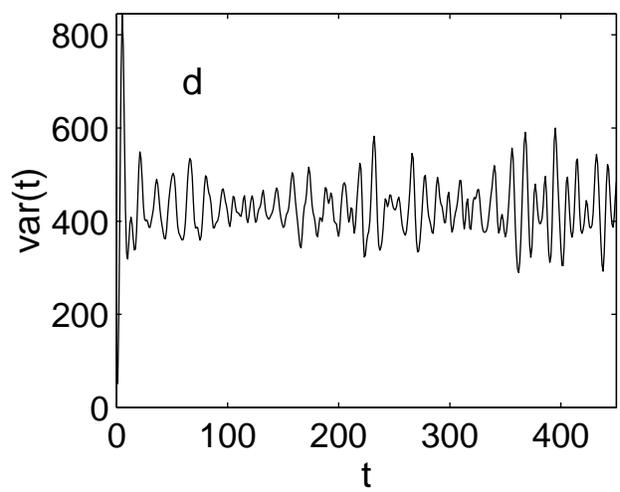
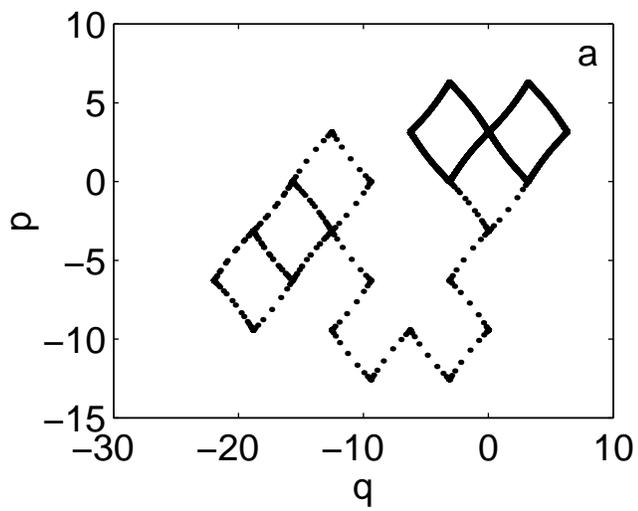
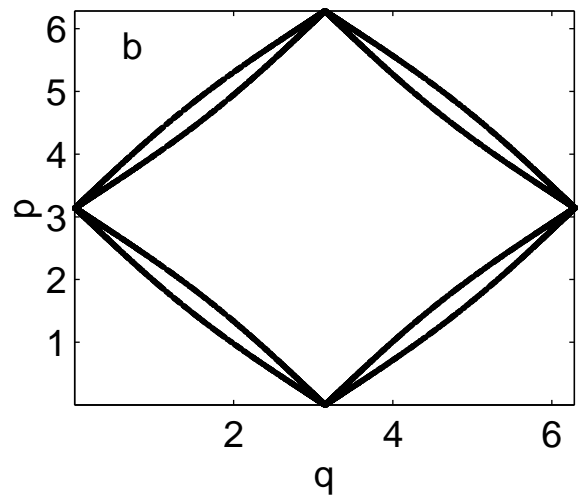
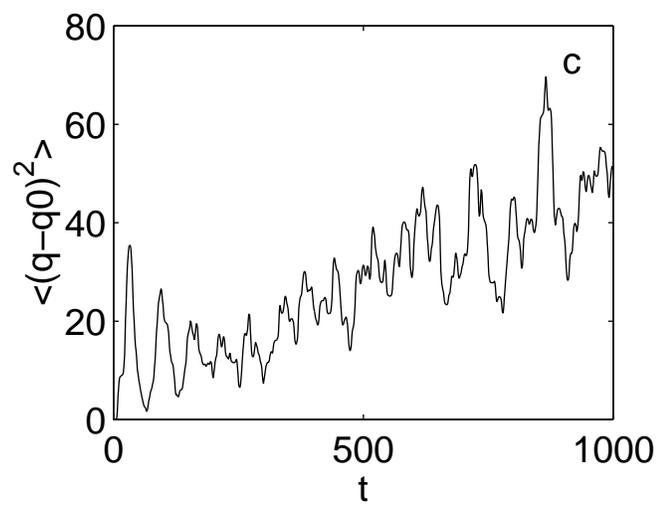
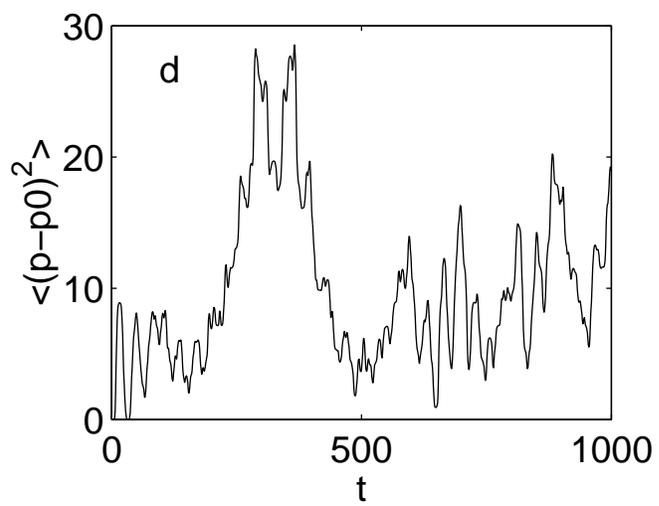
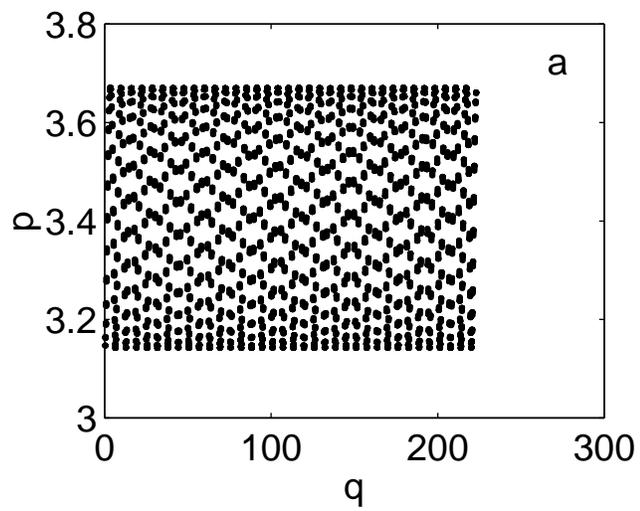
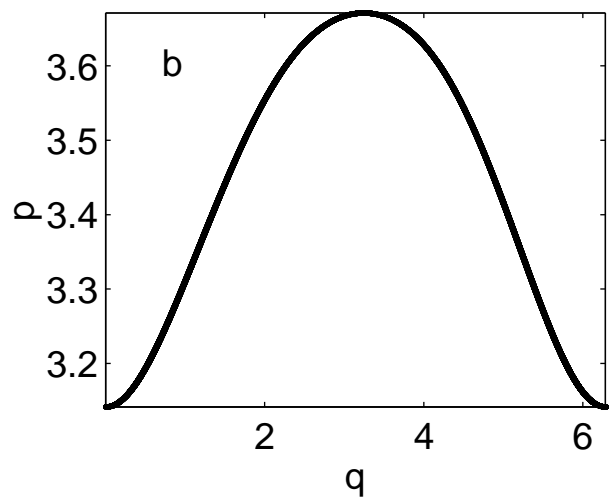
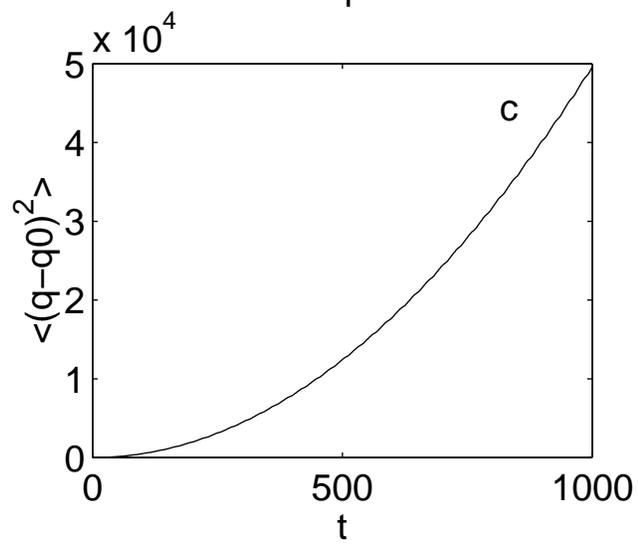
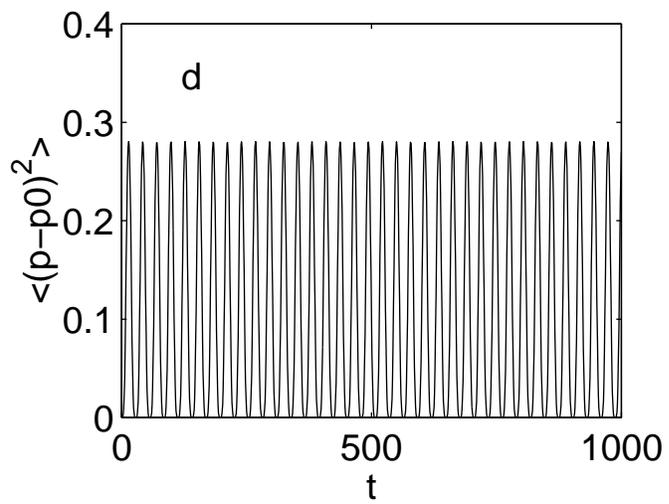
\end{document}